\documentclass[12pt]{article}

\usepackage{scicite}

\usepackage{times}

\usepackage{epsfig}

\newenvironment{sciabstract}{%
\begin{quote} \bf}
{\end{quote}}

\newcommand{\mdot}{\mbox{$\dot{M}$}}
\newcounter{lastnote}

\title{Why do millisecond pulsars have weaker magnetic fields compared
to ordinary pulsars?}

\author{Arnab Rai Choudhuri,$^{1}$ Sushan Konar$^{2\ast}$ \\
\normalsize{$^{1}$Department of Physics, Indian Institute of Science, 
Bangalore 560012, India,} \\ 
\normalsize{$^{2}$CTS \& Physics, Indian Institute 
of Technology, Kharagpur 721302, India} \\
\normalsize{E-mail: arnab@physics.iisc.ernet.in, sushan@cts.iitkgp.ernet.in}
}

\date{}

\begin{document} 


\baselineskip24pt


\maketitle 

\begin{sciabstract}
Millisecond pulsars, with magnetic fields weaker by three to four
orders compared to those of ordinary pulsars, are presumed to be 
neutron stars spun up by binary accretion. We expect the magnetic 
field to get screened by the accreted material. Our simulation of 
this screening mechanism shows, for the first time, that the field 
decreases by a purely geometric factor $\sin^{-7/2} \theta_{\rm P,i}$ 
before freezing to an asymptotic value, where $ \theta_{\rm P,i}$ is 
the initial angular width of the polar cap.  If $ \theta_{\rm P,i}$ 
lies in the range $5^o$--$10^o$, then the magnetic field diminution 
factor turns out to be $\sim 10^3$--$10^4$ in conformity with 
observational data.
\end{sciabstract}


Pulsars---which are rotating neutron stars---are seats of strongest 
magnetic fields known to mankind.  Magnetic fields of most pulsars 
are around $10^{11}$--$10^{13}$ G, whereas the typical rotation periods 
are about 1 s. However, there exist a handful of known pulsars with 
considerably smaller rotation periods, which also have much weaker 
magnetic fields around $10^8$ G. These so-called millisecond pulsars 
are often found in binary systems \cite{lorm01}.

It is believed that most of the neutron stars (barring possible 
{\it magnetars}\cite{kouv98}) are born with characteristics typical 
of ordinary pulsars. If the neutron star happens to be in a binary 
system, then it is possible for it to accrete matter with angular momentum 
from the binary companion. Millisecond pulsars are thought to be neutron 
stars which have been spun up in such a binary accretion process \cite{db99}.  
Since their magnetic fields are weaker by a factor $10^3$--$10^4$ compared 
to the magnetic fields of ordinary pulsars, presumably the magnetic field 
of the neutron star also decreases during the accretion phase. Several 
alternative scenarios have been proposed to explain how this decrease of 
magnetic field takes place, starting from models relating the field evolution 
to the spin evolution to models based on accretion-induced Ohmic decay in the 
heated crustal layers \cite{miri94,miri96,bd96,gepp94,urpn95,kb97,kb99a}. One 
possibility is that the magnetic field gets buried under the accreted matter.  
Although this idea has been around for a while 
\cite{bisn74,blnd79,taam86,roma90,roma95,cumm01,meph01}, we have, for the first 
time, carried out a detailed 2-D simulation to test this idea. Details of our 
calculations are presented elsewhere \cite{ck02,ck03}. One of the attractive 
features of this scenario is that the factor $10^3$--$10^4$ by which the magnetic 
field decreases can be explained in a very elegant and simple way as arising 
purely out of geometric considerations.

The strong magnetic field of an accreting neutron star channelizes the 
accreting material to flow through the polar regions.  As the magnetic 
field of the neutron star decreases because of the screening due to 
accreting material, it is less able to channelize the accretion flow 
and thereby the polar cap widens.  One can easily find out how the 
angular width $\theta_{\rm P}$ of the polar cap depends on the surface 
magnetic field $B_{\rm s}$ of the neutron star (see, for example, Shapiro 
and Teukolsky \cite{shap}). The field line starting from $\theta_{\rm P}$ 
at the surface of the neutron star, with a radius $r_{\rm s}$, is the last 
closed field line of the dipolar field and passes through the Alfv\'en 
radius $r_{\rm A}$. It easily follows that
$$
\sin \theta_{\rm P} = \left(\frac{r_{\rm s}}{r_{\rm A}} \right)^{1/2}. \eqno(1)
$$
Assuming that the ram pressure of the freely in-falling accreting material
at the Alfv\'en radius equals the magnetic pressure, a few steps of
easy algebra give 
$$
r_{\rm A} = (2GM)^{-1/7} r_{\rm s}^{12/7} B_{\rm s}^{4/7} {\mdot}^{-2/7}, \eqno(2)
$$
where $M$ is the mass of the neutron star and $\mdot$ the accretion rate.
It follows from (1) and (2) that
$$
\sin \theta_{\rm P} \propto B_{\rm s}^{-2/7}. \eqno(3) 
\label{eq_tht1}
$$
This is how the polar cap widens with the weakening magnetic field until
$\theta_{\rm P}$ becomes equal to $90^o$ when (3) obviously ceases to hold.
On taking $M = 10^{33}$ gm, $\mdot = 10^{-8} M_{\odot}$ yr$^{-1}$, 
$r_{\rm s} = 10$ km, $B_{\rm s} = 10^{12}$ G, we find from (2) that 
$r_{\rm A} \approx 300$ km.  Substituting this in (1), we conclude that 
the initial polar cap angle is of order $10^o$.   

The accreting materials falling through the two polar caps flow horizontally 
towards the equator in both the hemispheres.  At the equator, the 
opposing materials flowing in from the two poles meet, sink underneath the 
surface (inducing a counter-flow underneath the equator-ward flow at 
the surface) and eventually settle radially on the neutron star core.
With a suitably specified flow having these characteristics, we have 
studied kinematically how the magnetic field evolves with time, taking 
into account the fact that the polar cap width changes with the evolution 
of the magnetic field, thereby altering the velocity field also. Fig.~1 
shows the velocity field at an early stage (A) and at a late stage (B). 
We find that the equator-ward flow near the surface seen in Fig.~1A is quite 
efficient in burying the magnetic field underneath the surface. However, when 
the polar cap opens to $90^o$, the accretion becomes spherical and radial, as 
seen in Fig.~1B. It is found that such accretion is not efficient in burying 
the magnetic field any further. Fig.~2 shows magnetic field lines at different 
stages of evolution, whereas Fig.~3 plots the surface magnetic field at $45^0$ 
as a function of time.  Clearly, the magnetic field at the surface of the neutron 
star keeps decreasing until the polar cap opens to $90^o$, after which the 
magnetic field is essentially frozen, since the radial accretion cannot 
screen it any further. If $\theta_{\rm P,i}$ is the initial polar cap 
width, then it follows from (3) that the magnetic field would decrease 
by a factor $(\sin 90^o/ \sin \theta_{\rm P,i})^{7/2}$ from its initial
value before it is frozen to an asymptotic value.  On taking $\theta_{\rm P,i}$
in the range $5^o$--$10^o$, this factor turns out to be about $10^3$--$10^4$,
exactly the factor by which the magnetic fields of millisecond pulsars 
are weaker compared to the magnetic fields of ordinary pulsars.
 
Put another way, the magnetic field freezes when the Alfv\'en radius
becomes equal to the neutron star radius.  The asymptotic value of the 
surface magnetic field can be found directly from (2) by setting $r_{\rm A}$ 
equal to $r_{\rm s}$, which gives
$$B_{\rm asymp} = (2 G M)^{1/4} \mdot^{1/2} r_{\rm s}^{-5/4}. \eqno(4)$$
On using the various standard values mentioned before, we
find $B_{\rm asymp} \approx 10^8$ G. When the magnetic field falls to
this value, it can no longer channelize the accretion flow, resulting in
the flow becoming isotropic.  Such a flow is unable to screen the 
magnetic field any further.  After the accretion phase is over, the neutron 
star appears as a millisecond pulsar with this magnetic field. We propose 
this as the reason why 
millisecond pulsars are found with magnetic fields of order $10^8$ G. 

We would like to thank Somnath Bharadwaj and Sayan Kar for their
helpful comments.

\bibliography{tap_science.bib}

\bibliographystyle{Science}

\clearpage

\noindent {\bf Fig.\ 1.} Velocity fields induced inside 
the neutron star due to accreting
material settling on the surface.  The arrows indicate the flow 
velocities, whereas the dashed lines are contours of constant
$\nabla . (\rho {\bf v})$.  The velocity field is specified
in the code in such a way that it keeps changing with the evolution
of the magnetic field.  The details of how we do this are provided
elsewhere \cite{ck03}.  Here we show the velocity fields at two different
instants: (A) when the magnetic field is strong and the polar cap
angle is small (at a relatively early stage); and (B) when the
magnetic field has become much weaker and the polar cap has opened up 
(at a relatively late stage).  The dashed lines (contours of constant
$\nabla . (\rho {\bf v})$ ) indicate regions which are sources of
new material due to accretion.  We see in (A) that the new material is 
dumped in a narrow polar cap, inducing an equator-ward flow just
below the surface.  After reaching the equator where this flow
meets the oppositely-directed flow from the other pole, the flow
sinks underneath the surface, induces a counter-flow and eventually
settles on the core of the neutron star.  On the other hand, we
see in (B) that the new material falls isotropically all over the
surface and the induced flow is radially inward.  The horizontal
flow shown in (A) is expected in reality to be confined only within
1\% of the neutron star radius immediately below its surface \cite{ck02}.  
Here we have inflated that layer to 10\% of radius for easy visualization. \\

\noindent {\bf Fig.\ 2.} Magnetic field lines during different phases of 
evolution.  The dashed lines indicate the contours of constant $\nabla .
(\rho {\bf v})$ at the same instants. The topmost panel shows the 
initial magnetic field, which is evolved by solving the induction equation 
numerically, with the specified velocity field which keeps changing as 
the magnetic field weakens. (The velocity fields at an early stage and at 
a late stage are shown in Fig.\ 1.)  The details of the numerical code 
are given in the Appendix of our earlier paper \cite{ck02}, which also 
discusses the various boundary conditions used. The code has been adopted 
from a code which was used extensively for studying the evolution of the 
solar magnetic fields \cite{dikp94,nand02}. \\

\noindent {\bf Fig.\ 3.} We show how the magnetic field at the surface
at $45^0$ latitude changes with time (solid line).  The evolution
of the polar cap angle $\theta_P$ with time is also indicated
(dashed line).  It is clear the magnetic field decreases rapidly
till the polar cap opens to about $90^0$, after which the decay of the
magnetic field is significantly halted.

\clearpage

\begin{figure}
\begin{center}{\mbox{\epsfig{file=fig01.ps,width=200pt,angle=-90}}}\end{center}
\caption[]{}
\end{figure}

\begin{figure}
\begin{center}{\mbox{\epsfig{file=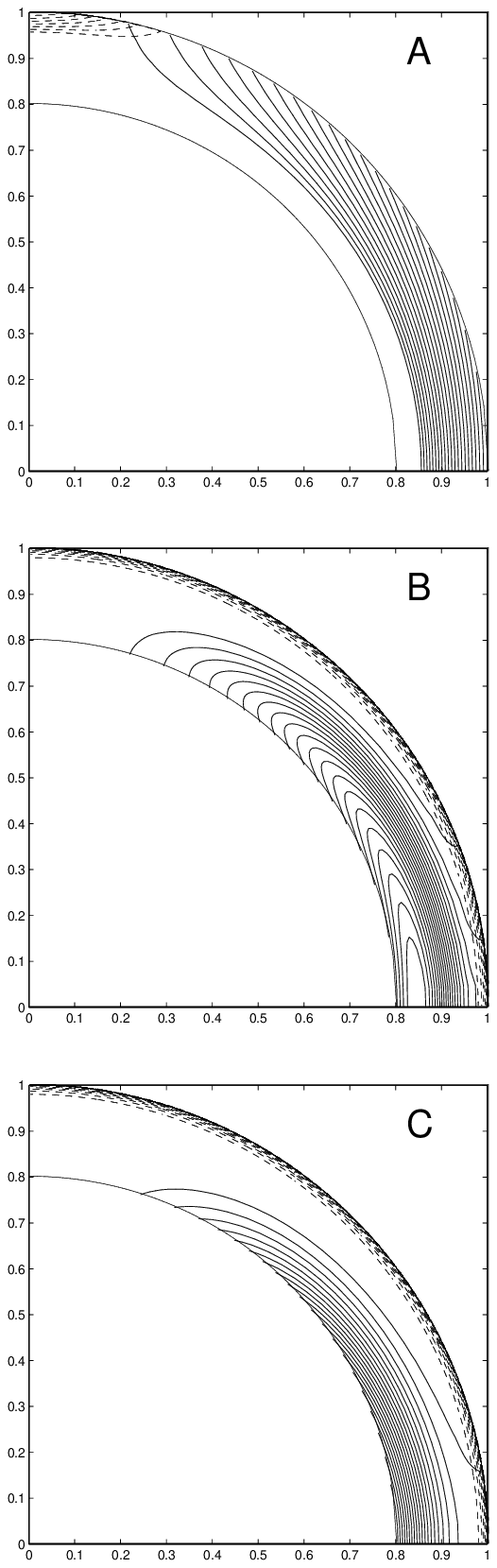,width=150pt}}}\end{center}
\caption[]{}
\end{figure}

\begin{figure}
\begin{center}{\mbox{\epsfig{file=fig03.ps,width=200pt,angle=-90}}}\end{center}
\caption[]{}
\end{figure}

\end{document}